\documentclass[prb,aps,superscriptaddress,twocolumn,longbibliography]{revtex4-2}
\usepackage{hyperref}
\usepackage{graphicx}
\usepackage{bm}
\usepackage{amsmath}
\usepackage{amsfonts}
\usepackage{amssymb}
\usepackage{epsfig}
\usepackage{color}
\usepackage{ulem}

\begin{document}

\title{Diode effect in the Fraunhofer pattern of disordered planar Josephson junctions}

\author{Luca Chirolli}
\email{luca.chirolli@unifi.it}
\affiliation{Department of Physics and Astronomy, University of Florence, I-50019 Sesto Fiorentino, Italy}
\affiliation{Quantum Research Center, Technology Innovation Institute, Abu Dhabi, UAE}

\author{Angelo Greco}
\affiliation{NEST, Istituto Nanoscienze-CNR and Scuola Normale Superiore, I-56127 Pisa, Italy}

\author{Alessandro Crippa}
\affiliation{NEST, Istituto Nanoscienze-CNR and Scuola Normale Superiore, I-56127 Pisa, Italy}

\author{Elia Strambini}
\affiliation{NEST, Istituto Nanoscienze-CNR and Scuola Normale Superiore, I-56127 Pisa, Italy}

\author{Mario Cuoco}
\affiliation{CNR-SPIN c/o Universita' di Salerno, Fisciano, Italy}

\author{Luigi Amico}
\affiliation{Quantum Research Center, Technology Innovation Institute, Abu Dhabi, UAE}
\affiliation{Dipartimento di Fisica e Astronomia 'Ettore Majorana', Via S. Sofia 64, 95123 Catania, Italy}
\affiliation{INFN-Sezione di Catania, Via S. Sofia 64, 95127 Catania, Italy}

\author{Francesco Giazotto}
\affiliation{NEST, Istituto Nanoscienze-CNR and Scuola Normale Superiore, I-56127 Pisa, Italy}

\begin{abstract}
{\bf Abstract.---}The Josephson diode effect describes the property of a Josephson junction to have different values of the critical current for different direction of applied bias current and it is the focus of intense research thanks to the possible applications. The ubiquity of the effect experimentally reported calls for a study of the impact that disorder can have in the appearance of the effect. We study the Fraunhofer pattern of planar Josephson junctions in presence of different kinds of disorder and imperfections and we find that a junction that is {\it mirror} symmetric at zero-field forbids the diode effect and that the diode effect is typically magnified at the nodal points of the Fraunhofer pattern. The work presents a comprehensive treatment of the role of pure spatial inhomogeneity in the emergence of a diode effect in planar junctions, with an extension to the multi-terminal case and to systems of Josephson junctions connected in parallel.
\end{abstract}

\maketitle

\section{Introduction}

Non-reciprocal effects in superconductors manifest with the magnitude of the critical current acquiring a dependence on the direction of the applied current. Such a feature results in the so-called Superconducting Diode Effect (SDE), and ultimately yields supercurrent rectification. Motivated both by fundamental questions about the underlying generating mechanisms and by potential technological applications in dissipationless nanoelectronics \cite{upadhyay2024}, a large amount of scientific work has recently focused on the diode effect \cite{ando2020,baumgartner2022,bauriedl2022,strambini2022,costa2023,nadeem2023}. The latter has been observed in the critical current of two classes of systems: {\it i)} in bulk materials, where it is a thermodynamic property of the superconducting state, in which case the term JDE is more appropriate, and {\it ii)} in Josephson junctions, where the switching to a dissipative state involves mostly the junction and the bulky terminals maintain their superconducting properties. In this case, we refer to the Josephson Diode Effect (JDE).

Nonreciprocal critical supercurrents have been measured in noncentrosymmetric systems with Rashba spin-orbit coupling and in-plane magnetic field \cite{ando2020,wakatsuki2017,zhang2020}, in the fluctuation state of polar superconductors \cite{itahashi2020}, in nanopatterned devices \cite{lyu2021}, non-centrosymmetric superconductor/ferromagnet multilayers \cite{narita2022}, and in trilayer twisted graphene in zero magnetic fields, where it points to a time-reversal symmetry breaking state \cite{lin2022}. SDE has been proposed and measured for finite-momentum Cooper pairs \cite{lin2022,noah2022,pal2022,peihao2025}, and it has been measured in twisted high $T_c$ superconductors junctions \cite{frankzhao2023,ghosh2024}.

In Josephson systems, JDE has been proposed and measured in a number of different devices, such as junctions with conventional superconductors and central regions with spin-orbit interactions \cite{baumgartner2022,costa2023,lotfizadeh2024}, Dayem bridges \cite{souto2022,MertBorzkurt2023,margineda2023sign}, and interferometric setups featuring a high-harmonic content \cite{greco2023,greco2024}. JDE has been observed in junctions hosting screening currents \cite{hou2023,sundaresh2023}, self-field effects induced by supercurrents \cite{krasnov1997,golod2022}, and phase reconfiguration induced by Josephson current and kinetic inductance \cite{chen2024}. It has also been observed in multi-terminal devices \cite{gupta2023,zhang2024}  gate-tunable junctions  \cite{ciaccia2023,gupta2023}, and through the back-action mechanism \cite{margineda2025back,de2024quasi}.

It is widely accepted that the crucial conditions for the emergence of the diode effect are time-reversal and spatial-inversion symmetry breaking, together with high transparency in Josephson junction realizations. However, whether inversion and time-reversal symmetries are broken in a global intrinsic way or at a local extrinsic level and how this impacts the experimental phenomenology remains unclear. In particular, the role of disorder in the junction, either via short-range impurities, or through geometric asymmetries and applied gate voltages, is not yet evident.  Within this context, an analysis of the symmetries and their impact on the diode effect motivates the present work, especially concerning the role of spatially dependent phase patterns in determining a reciprocity symmetry breaking. To this end, interferometry is a powerful investigation tool. 

\begin{figure}[b]
\includegraphics[width=1.0\linewidth]{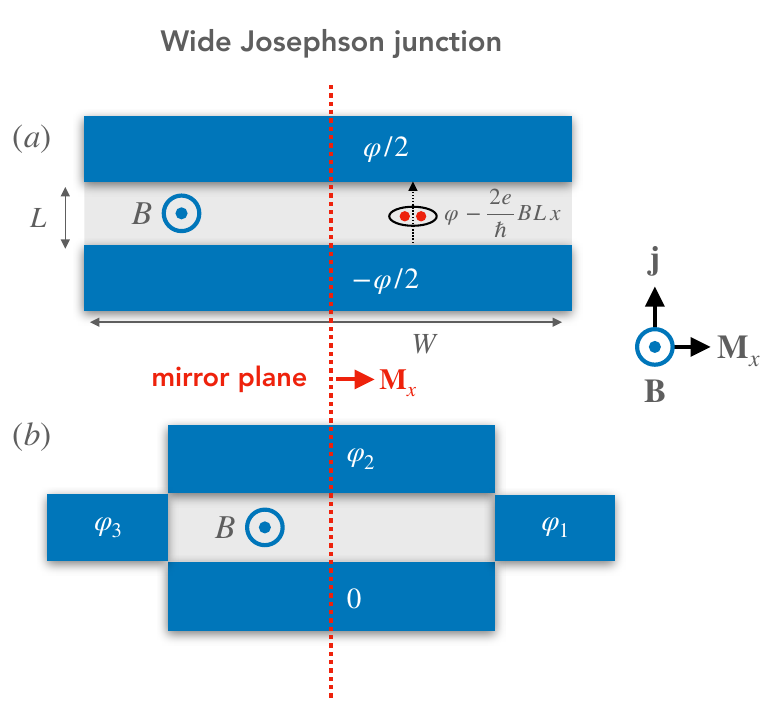}
    \caption{\label{Fig0}{\bf Schematics of a planar Josephson junction.} A magnetic field yields a position-dependent phase that induces a diffractive Fraunhofer pattern in the critical current. The cross-product between the current and the magnetic field determines the mirror symmetry plane. (a) two-terminal junction and (b) four-terminal junction. }     
\end{figure}

In this work, we investigate the Fraunhofer diffraction pattern in planar Josephson junctions focusing on the role of spatial inhomogeneity in the emergence of a diode effect at finite flux. In a planar Josephson junction, schematically shown in Fig.~\ref{Fig0}(a), two superconducting terminals are coupled via a central normal region threaded by a magnetic flux that acts as a diffraction center for Cooper pairs. In traversing the junction in the direction of the phase bias $\varphi$, Cooper pairs acquire a phase that spatially depends on the transverse direction, 
\begin{equation}
\varphi(x)= \varphi-2\pi \Phi(x)/\Phi_0, 
\end{equation}
with $\Phi(x)$ the flux of the field through the area to the left of $x$ and $\Phi_0=h/2e$ the quantum superconducting flux, which yields the typical interferometric pattern in the critical current. It becomes clear that such a device is susceptible to spatial inhomogeneity, which locally affects the phase acquired by the Cooper pairs and the resulting interferometric pattern. As one of the most important results, we find that, in junctions that in absence of magnetic field are symmetric under a specific mirror symmetry, a JDE can be ruled out. In turn, in all the devices considered in which the relevant mirror symmetry is broken a JDE always occurs. 

By means of the rectification coefficient as a figure of merit, we analyze the Fraunhofer pattern of different specific configurations that break mirror symmetry, such as smoothly varying profiles, structures with short-range disorder, and geometrically asymmetric devices. We report qualitatively similar results, with spikes in correspondence of the nodes of the Fraunhofer pattern, where JDE is magnified due to destructive interference, that suppresses the current in one direction more than in the opposite direction. A statistical analysis of the rectification coefficient in disordered systems shows a common trend of its root mean square, that can be ascribed to mesoscopic critical current fluctuations and that suggest further investigation. The analysis is extended to arrays of Josephson junctions connected in parallel and to multiterminal junctions, where the phase at different terminals can be used as a knob to modulate the Fraunhofer pattern and tune the diode effect at finite magnetic flux, thus extending the results on multiterminal JDE \cite{gupta2023,zhang2024,huamani2024} to the case of the Fraunhofer interference pattern.

The present work offers a comprehensive analysis of the role of mirror symmetry and of disorder in the emergence of the Josephson diode effect in realistic devices.

\section{Results}

The Josephson diode effect in a junction is studied by assessing the equilibrium current $I(\varphi)$ as a function of phase bias $\varphi$. The latter can be expressed using the free energy of the system  $F(\varphi)$ at temperature $T$ as $I(\varphi)=(2\pi/\Phi_0)\partial F(\varphi)/\partial \varphi$ \cite{bardeen1969}. The critical currents for positive- and negative-bias currents are defined as
\begin{eqnarray}
    I_{c}^+=\max_{\varphi} I(\varphi),\qquad
    I_{c}^-=\min_{\varphi} I(\varphi),
\end{eqnarray}
and a JDE occurs when $I^+_c\neq |I^-_c|$. The efficiency of the diode effect is quantified by the rectification coefficient
\begin{equation}\label{Eq:Eta-def}
    \eta=\frac{I_c^+-|I_c^-|}{I_c^++|I_c^-|}.
\end{equation}
We are interested in the dependence of the critical currents on the external flux $\Phi$ produced by an orbital magnetic field that threads the junction. From the Onsager reciprocity relations, it follows that, in general, we have $I_c^+(\Phi)=-I_c^-(-\Phi)$. In particular, we now demonstrate that for a system that is mirror-symmetric in zero field we have $I_c^+(\Phi)=-I_c^-(\Phi)$.

\subsection{Absence of JDE in mirror-symmetric junctions}

We formalize this concept by focusing on the role of spatial mirror symmetry in the junction using a scattering matrix approach \cite{beenakker1991,beenakker1992,beenakker1997,vanheck2014}. The latter is particularly suited to study mesoscopic coherent systems, allowing us to make general statements, independently of the particular realization. By knowing the scattering matrix $s(\epsilon)$ describing the central normal region and assuming its dependence on energy $\epsilon$ can be neglected on the scale of the gap $\Delta$, we can obtain the Andreev spectrum by the singular values of the complex matrix \cite{vanheck2014} (see also \footnote{Supplementary Material: Sec. I - Scattering Matrix Formulation})
\begin{equation}
A=\frac{1}{2}(r_{\rm A}s+s^Tr_{\rm A}),
\end{equation}
with $r_A={\rm diag}(e^{-i\varphi/2}\openone_L,e^{i\varphi/2}\openone_R)$ and $\openone_{L(R)}$ are identity matrices with the dimension of the number of open transport channels in the $L(R)$ lead.  

We consider a specific mirror symmetry of the junction. With reference to Fig.~\ref{Fig0}(a), this is given by $M_x$, such that $M_x x M_x^{-1}=-x$. For a general scattering matrix $s$ that is mirror-symmetric in the absence of the external field, we can write
\begin{equation}
    M_x s(\Phi) M_x^{-1}=s(-\Phi),
\end{equation}
with $\Phi$ the total flux through the junction. Furthermore, by the Onsager reciprocity relations we know that $s^T(\Phi)=s(-\Phi)$, from which it follows that the matrix $A$ transforms under mirror $M_x$ as
\begin{eqnarray}
    M_x A(\varphi,\Phi) M_x^{-1} = r_A A(-\varphi,\Phi) r_A.
\end{eqnarray}
This results from the fact that the phase difference $\varphi$ is unaltered by the mirror symmetry $M_x$. Since the singular values of $A$ give the Andreev spectrum, a similarity between $A(\varphi,\Phi)$ and $A(-\varphi,\Phi)$ guarantees that the free energy satisfies $F(\varphi,\Phi)=F(-\varphi,\Phi)$, from which it follows that the current satisfies
\begin{equation}
    I(\varphi,\Phi)=-I(-\varphi,\Phi),
\end{equation}
and from the Onsager reciprocity relations a diode effect cannot occur. Without zero-field mirror symmetry, a diode effect cannot be ruled out, and its features depend on the system microscopic or macroscopic spatial details.  

It is crucial to appreciate how other choices of zero-field mirror symmetry cannot yield conclusive statements about the absence of JDE. For example, for a system whose scattering matrix satisfies $M_ys(\Phi)M_y^{-1}=s(-\Phi)$, we obtain $M_yA(\varphi,\Phi)M_y^{-1}=r_A^*A(\varphi,\Phi)r_A^*$, so that we cannot conclude anything regarding the presence or absence of JDE. Furthermore, it is interesting to consider the inversion symmetry, or parity, $\hat{P}$, which maps ${\bf r}\to -{\bf r}$. Inversion symmetry can be decomposed as the product of three orthogonal mirror symmetries, $\hat{P}=M_xM_yM_z$. Assuming that $M_y$ is broken implies breaking of the inversion symmetry. At the same time, a zero-field mirror symmetry about $M_x$ rules out the JDE, showing that, generally, {\it inversion symmetry breaking is only a necessary condition}. 

Introducing ${\bf n}$ as the vector normal to the plane defining a given mirror transformation $M_{\bf n}$, we have that JDE appears when the system breaks the mirror for which ${\bf n}$ is parallel to the cross product between the applied external field ${\bf B}$ and the direction of the current bias ${\bf j}$ yielding the junction phase difference $\varphi$, 
\begin{equation}
{\bf n}\parallel {\bf j}\times {\bf B},
\end{equation}
already in absence of the magnetic field. With reference to Fig.~\ref{Fig0}(a), we have ${\bf n}=\hat{\bf x}$. This result is general and, in particular, it can be extended to the case of SDE and JDE in systems with Rashba spin-orbit interaction and in-plane Zeeman field \cite{baumgartner2022,costa2023,nadeem2023}: in that case the magnetic field is in the plane of the junction, orthogonal to the current bias, and the Rashba spin-orbit term appears due to the breaking of the mirror symmetry about the plane of the junction. We point out that a magnetic field alone is a source of mirror symmetry breaking for any mirror plane containing the field, as can be seen by choosing a gauge for the vector potential ${\bf A}({\bf r})=(0,Bx,0)$. At the same time, if the magnetic field is the only source of mirror symmetry breaking and the field is uniform, the junction spatial symmetry dictates how the scattering matrix transforms under mirror symmetry. 

Although a symmetry analysis cannot guarantee the onset of JDE, an important point is that junctions that at zero-field break the $M_x$ symmetry always show JDE. In the next sections, an analysis of several different kinds of structures that break the mirror symmetry $M_x$ in the zero field will corroborate such a statement.

\begin{figure*}[t]
	\centering
    \includegraphics[width=1.0\linewidth]{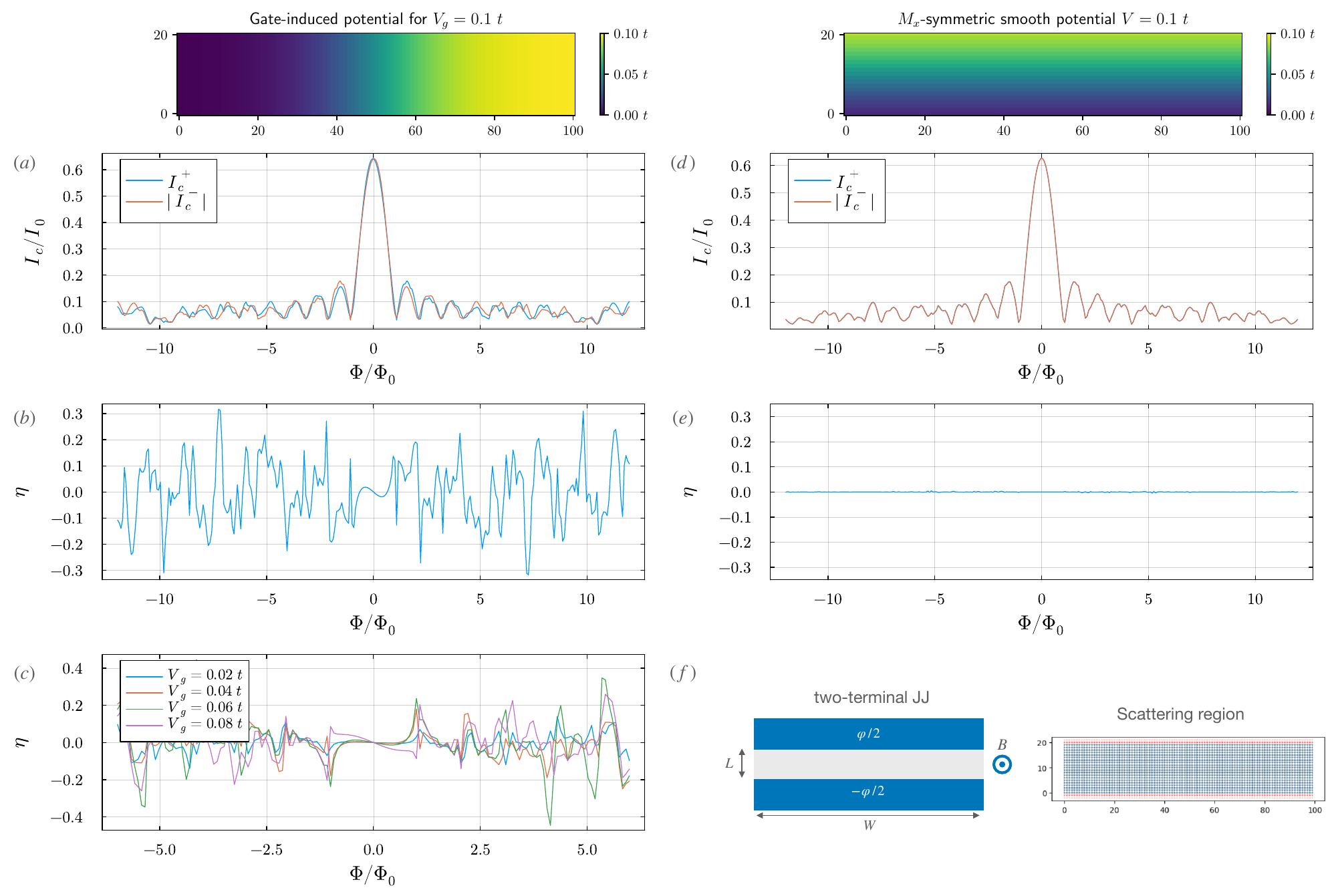}
	\caption{{\bf JDE in the Fraunhofer pattern for a planar Josephson junction with long-wavelength spatial modulations}. The junction is schematically depicted in (f), in which two superconducting terminals kept at phase different $\varphi$ are contacted to a normal region of width $W$, length $L$, and pierced by a magnetic field $B$ that induces a flux $\Phi=B A$, with $A=WL$ the area of the central normal region. (a)  Critical currents $I_c^\pm$ as a function of the external flux $\Phi$ resulting from the local onsite potential $\delta U_i=V_g(\tanh((x_i-W/2)/W_c)+1)/2$, for $V_g=0.1~t$, $\mu=0.3~t$, and $W_c=20~a$, shown in the top-left of the figure. (b) Rectification coefficient $\eta$ for the critical currents shown in (a). (c) Dependence of the rectification coefficient $\eta$ of (b) on the gate potential $V_g$, for different values of $V_g$. (d) Critical currents $I_c^\pm$ as a function of the external flux $\Phi$ resulting from the local onsite potential $\delta U_i=V(\tanh((y_i-L/2)/W_c)+1)/2$, shown in the top-right of the figure.  (e) Rectification coefficient $\eta$ for the critical currents shown in (d). The residual value arises due to the discretization of the grid. (f) Schematics of the wide Josephson junction and square lattice of lateral dimensions $W=100 ~a$ and $L=20~a$, in terms of a microscopic unit length $a$, employed in the tight-binding model to calculate the scattering matrix. Semi-infinite leads are attached to the scattering region's top and bottom. A potential barrier of strength $U_{\rm pot}=0.2~t$ is added at the interface with the leads.
 \label{Fig2}}    
\end{figure*}

\subsection{Planar Josephson junctions}
\label{Sec:WideJunction}

Wide Josephson junctions comprise an extended normal central region connected to two wide superconducting leads, schematically depicted in Fig.~\ref{Fig0}(a). Here,  we focus on two-terminal planar systems and consider three instances of spatial inhomogeneity: {\it i)} a geometrically inversion symmetric ballistic junction with a spatial potential in the central area that varies on a long length scale, on order of the junction lateral sizes, {\it ii)} a clean ballistic system with a trapezoidal geometric shape, {\it iii)} a geometrically inversion symmetric, ballistic junction with a spatially varying potential that varies on a short length scale, on order of the interatomic distances. The case {\it i)} is compatible with the charge profile induced by the back and side gates, and the case {\it iii)} is compatible with local disorder.

We model the central region using a generic tight-binding model on a square lattice, as the one schematized in Fig.~2(f) of the main text, described by the Hamiltonian
\begin{equation}
    \label{Eq:H_tight_binding}
    H_c=\sum_{i,\sigma}U_ic^\dag_{i\sigma} c_{i\sigma}-t\sum_{<i,j>\sigma}e^{i\theta_{i,j}}c^\dag_{i\sigma}c_{j\sigma}+{\rm H.c.},
\end{equation}
where $c_{i\sigma}$ describes electrons with spin $\sigma=\uparrow,\downarrow$ at position ${\bf r}_i$, and the index $i$ runs over all the lattice points of the central region. The square lattice has a sublattice symmetry that yields a particle-hole symmetry in the spectrum. Electrons have onsite energy $U_i=4t-\mu+\delta U_i$, comprising a chemical potential contribution $\mu$ and an additional onsite potential $\delta U_i\equiv \delta U({\bf r}_i)$, and can hop among nearest-neighboring sites with amplitude $-te^{i\theta_{i,j}}$, where the Peierls phase $\theta_{i,j}=\frac{e}{\hbar}\int_{{\bf r}_i}^{{\bf r}_j}{\bf A}({\bf r})\cdot d{\bf s}$ is due to the external magnetic field ${\bf B}=\nabla\times {\bf A}$. The central region is attached to the superconducting leads, and potential barriers that shift local energy are added to the contact interface. We consider transport at zero energy in the limit in which the energy dependence of the scattering matrix can be neglected, that corresponds to a system length $L< \xi$, with $\xi$ the coherence length of the superconducting leads, and numerically simulate several systems employing the KWANT package \cite{groth2014} (see \footnote{Supplementary Material: Sec. II - Comparison with realistic systems}). 

As a general trend, we find that when the zero-field mirror symmetry of the junction is broken, a JDE appears, which is highly enhanced at the nodes of the Fraunhofer pattern, where the critical currents do not go to zero simultaneously.

\begin{figure*}[t]
	\centering
    \includegraphics[width=1.0\linewidth]{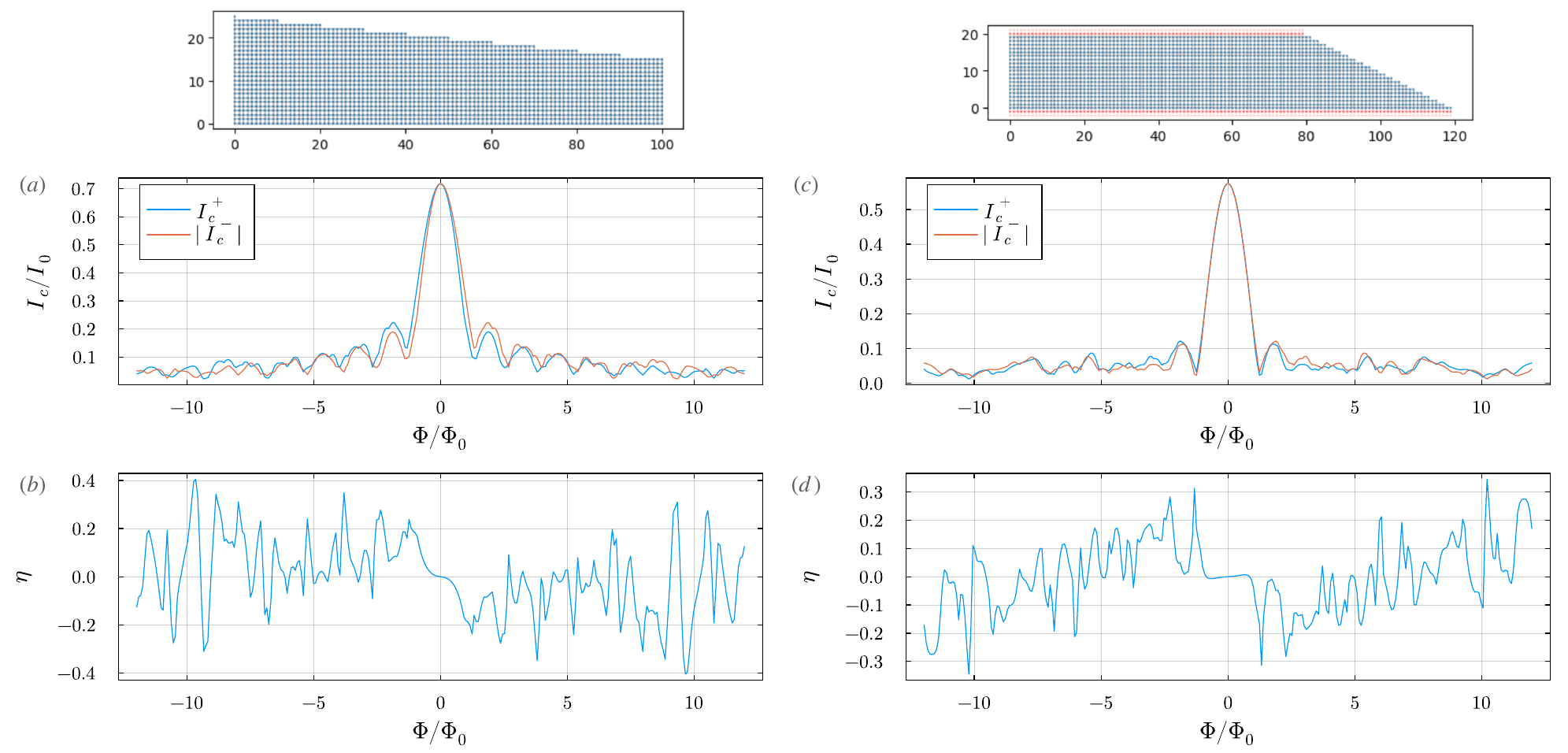}
	\caption{{\bf JDE in the Fraunhofer pattern of geometrically asymmetric junctions.} (a) Critical currents $I_c^\pm$ as a function of the external flux $\Phi$ for the trapezoidal junction shown in the top-right inset, with width $W=100~a$ and left and right lengths $L_l=25~a$, $L_r=15~a$. (b) Rectification coefficient $\eta$ for the critical currents shown in (a). (c) Critical currents $I_c^\pm$ as a function of the external flux $\Phi$ for the trapezoidal junction shown in the top-right inset, with bottom and top width $W_b=120~a$, $W_t=80~a$, and length $L=20~a$. (d) Rectification coefficient $\eta$ for the critical currents shown in (c). For all panels we set $\mu=0.3~t$, barrier $0.2~t$, and $\delta U_i=0$.  \label{Fig3}}     
\end{figure*}

\subsubsection{Smooth Potentials}

\begin{figure*}[t]
	\centering
    \includegraphics[width=1.0\linewidth]{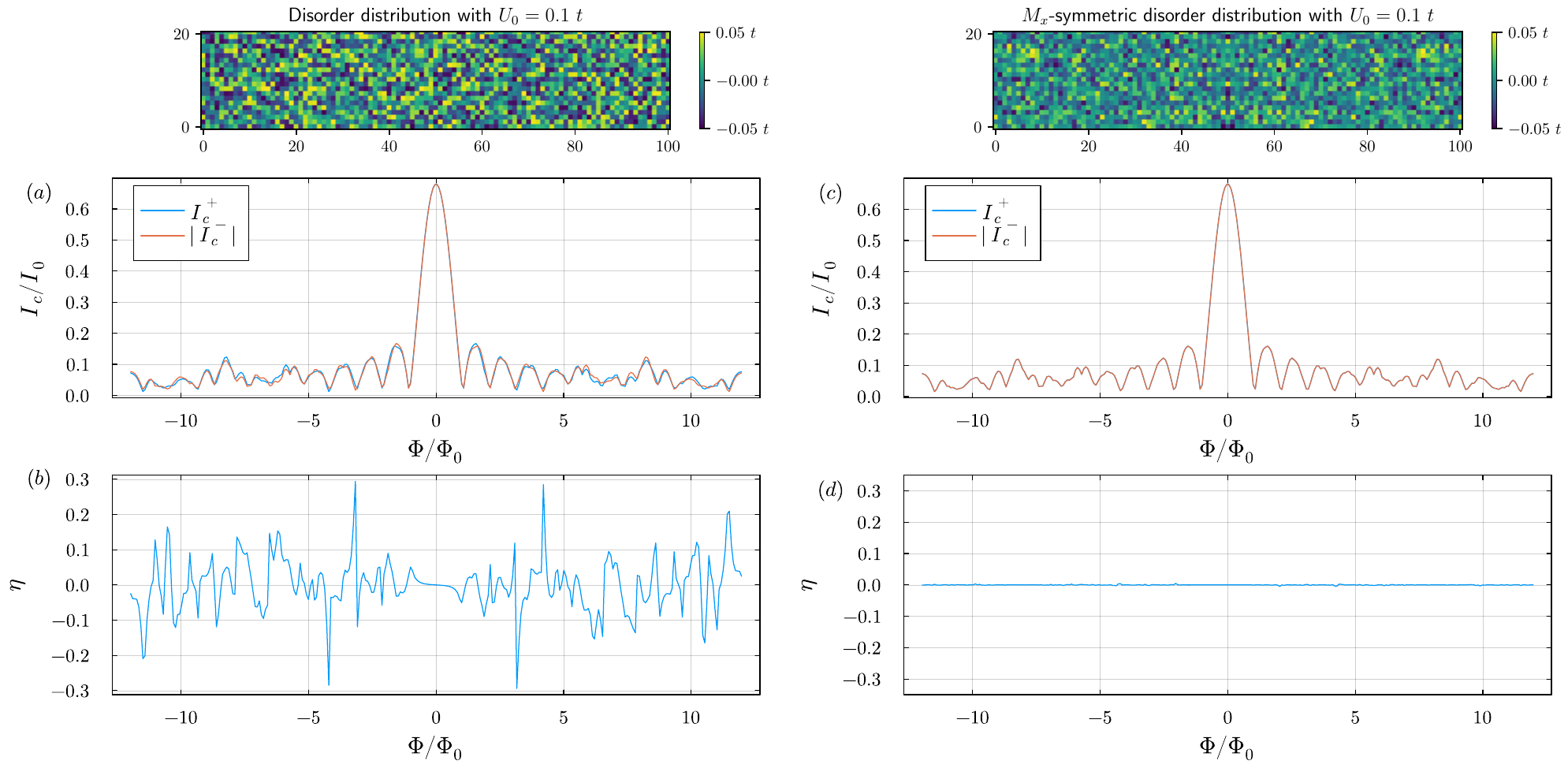}
	\caption{{\bf Fraunhofer interference pattern of a planar Josephson junction with short-range disorder.} The junction structure is as in Fig.~\ref{Fig2}. (a)  Critical currents $I_c^\pm$ as a function of the external flux $\Phi$ resulting from the local onsite potential $\delta U_i$, with $-U_0/2<\delta U_i<U_0/2$ randomly distributed for $U_0=0.1~t$, shown in the top-right inset.  (b) Rectification coefficient $\eta$ for the critical currents shown in (a). (c)  Critical currents $I_c^\pm$ as a function of the external flux $\Phi$ resulting from a local onsite potential $\delta U_i$, with $-U_0/2<\delta U_i<U_0/2$ randomly distributed for $U_0=0.1~t$ and symmetrized $\delta U_i\to (\delta U_i+M_x \delta U_i M_x^{-1})/2$, shown in the top-right inset.  (d) Rectification coefficient $\eta$ for the critical currents shown in (c). The chemical potential is set to $\mu=0.3~t$.
 \label{Fig4}}    
\end{figure*}

We first consider a central rectangular region of width $W=100~a$ and length $L=20~a$, in terms of a microscopic discretization length $a$, schematized in Fig.~\ref{Fig2}(f). We set the chemical potential at $\mu=0.3~t$ in both the central region and in the leads and apply a non-uniform potential $\delta U(x,y)=V_g(\tanh((x-W/2)/L_s)+1)/2$. The potential, which also comprises a shift of the chemical potential, is not symmetric under $M_x$, such that $M_x \delta U(x,y)M_x^{-1}\neq \delta U(-x,y)$, and it is shown in the upper right inset of Fig.~\ref{Fig2}(a) for a value of $V_g=0.1~t$, producing a smooth profile that changes on a length $L_s=20~a$. Potential barriers that shift the local energy by $0.2~t$ are added at the interface with the contacts. In Fig.~\ref{Fig2}(a), we show the associated Fraunhofer interference pattern for the critical currents $I_c^\pm$. We see that the presence of a smooth potential breaking the mirror symmetry $M_x$ produces a diode effect, as quantified by the rectification coefficient $\eta$ shown in Fig.~\ref{Fig2}(b), reaching rectification $30\%$. 

The fact that a smooth, long-wavelength potential can yield a diode effect suggests that the latter can be tuned by applying a gate. The potential shown in the inset of Fig.~\ref{Fig2}(a) well approximates the one produced by a back gate that occupies the right side of the junction. In Fig.~\ref{Fig2} (c), we plot the rectification efficiency for few values of the gate potential, $V_g/t=0.02, 0.04, 0.06, 0.08$. In the first lobe we clearly see a trend for which the rectification increases with the gate potential. In turn, we see no clear systematic evolution of the curves for increasing gate strength. The rectifications shows spikes in correspondence of the nodes of the Fraunhofer pattern, which are due destructive interference acting in different ways for trajectories that go from one lead to the other and vice versa. 

We then consider a potential that is even under $M_x$ but that is not symmetric under $M_y$, thus breaking the inversion symmetry $\hat{P}$. Specifically, we choose $\delta U(x,y)=V\tanh((y-L/2)/L_s)$. The potential satisfies $M_x \delta U(x,y)M_x^{-1}=\delta U(-x,y)$, $M_y \delta U(x,y)M_y^{-1}\neq\delta U(x,-y)$, and it is shown in the top-right inset of Fig.~\ref{Fig2}(d) for a value of $V_g=0.1~t$, yielding a smooth profile that changes on a length $L_s=20~a$. As clearly shown in Fig.~\ref{Fig2}(d), the two curves $I^\pm_c(\Phi)$ cannot be distinguished and the rectification coefficients are less than $0.4\%$,  as shown in Fig.~\ref{Fig2}(e). This result is significant, as it corroborates the finding that breaking of inversion symmetry cannot guarantee a JDE and that the presence of a symmetric profile under $M_x$ guarantees no JDE.

\subsubsection{Geometric asymmetry}

In Fig.~\ref{Fig3} we consider the case {\it ii)} of a clean system with a trapezoidal shape. We consider two possible cases, one in which the distance between the two leads varies with the lateral position, shown in the upper left panel of Fig.~\ref{Fig3} (a), and one in which the size of the leads is different, shown in the upper right panel i in Fig.~\ref{Fig3} (c). We see that the shape of the central region is sufficient to break the $M_x$ mirror symmetry of the junction, producing a diode effect, as quantified by the rectification coefficient $\eta$ shown in Figs.~\ref{Fig3} (b) and (d), reaching up to $40\%$. In Fig.~\ref{Fig3}(a), we see that the position-dependent distance between the leads for the entire structure yields a phase average that suppresses destructive interference. In contrast, in Fig.~\ref{Fig3} (c), the phase averaging arises from additional trajectories on the system's right side. In the present case, the system is fully ballistic and coherent, and the averaging effect is solely due to the geometry that introduces paths with different lengths that add up to build the interference pattern.

\subsubsection{Short wavelength disorder}

We then consider the case {\it iii)} of a potential that varies on a short length scale, with local impurities of strength $\delta U_i$ randomly distributed between $-U_0/2<\delta U_i<U_0/2$, as shown in the upper left panel of Fig.~\ref{Fig4} for $U_0=0.1~t$. This is a relatively weak short-wavelength disorder of the Anderson type, that yields a long mean-free path $\ell_{\rm mf}$, much larger than the Fermi wavelength. The resulting Fraunhofer interference pattern for the critical currents $I_c^\pm$ is shown in Fig.~\ref{Fig4}(a), together with the rectification coefficient in Fig.~\ref{Fig4}(b). Analogously to the previous case, the broken mirror symmetry $M_x$ is sufficient to give a diode effect in the Fraunhofer interference pattern, which is qualitatively very similar to the case of a smooth uniform potential of Fig.~\ref{Fig2}(a,b). In particular, spikes appear in the rectification in correspondence of the nodes of the Fraunhofer pattern that are accompanied by sign changes of the rectification. In addition, we check that by imposing symmetry under $M_x$ the JDE disappears. To this end, we first generate a configuration with $\delta U_i$ randomly distributed between $-U_0/2<\delta U_i<U_0/2$, and then symmetrize it such that $\delta U_i\to (\delta U_i+M_x \delta U_i M_x^{-1})/2$. The resulting potential is shown in the upper right panel  of Fig.~\ref{Fig4}(c) and the associated currents $I_c^\pm$ shown in Fig.~\ref{Fig3}(c) are not distinguishable with the naked eye, as can be checked in Fig.~\ref{Fig4}(d) where a maximal rectification of $0.4\%$ appears.

\subsection{Disorder and Fluctuations}

A clear result that appears in all the simulations is the strong fluctuation in the rectification coefficient as we vary the magnetic field. By construction, the rectification coefficient is sensitive to variation of the critical currents and fluctuations in the latter are clearly amplified in the JDE. In Ref.~\cite{beenakker1991,beenakker1992,beenakker1997}, fluctuations in the critical current have been assessed in Josephson junctions characterized by a central disordered normal region, in analogy to universal conductance fluctuations \cite{altshuler1985,LeeStone1985,imry1986}. There, it was shown that for systems in the regime $\ell_{\rm mf}, L< \xi$, in which neglecting the energy dependence of the scattering matrix is justified, fluctuations from sample to sample give a ${\rm rms}~I_c=I_0$, with $I_0=e\Delta/\hbar$.  This result follows from the fact that the current is a {\it linear statistics} \cite{imry1986,stone1991,altshuler2012mesoscopic}, so that ${\rm rms}~I_c/I_0={\cal O}(1)$ regardless of the number $N$ of open channels \cite{beenakker1992}, as long as $\ell_{\rm mf}\ll L \ll N\ell_{\rm mf}$, and the lateral size is not much larger than the longitudinal one. If the junction is much wider than longer, as in the present case of a planar junction, we have ${\rm rms}~I/I_0\propto \sqrt{W/L}$.

\begin{figure}[t]
	\centering
    \includegraphics[width=1.0\linewidth]{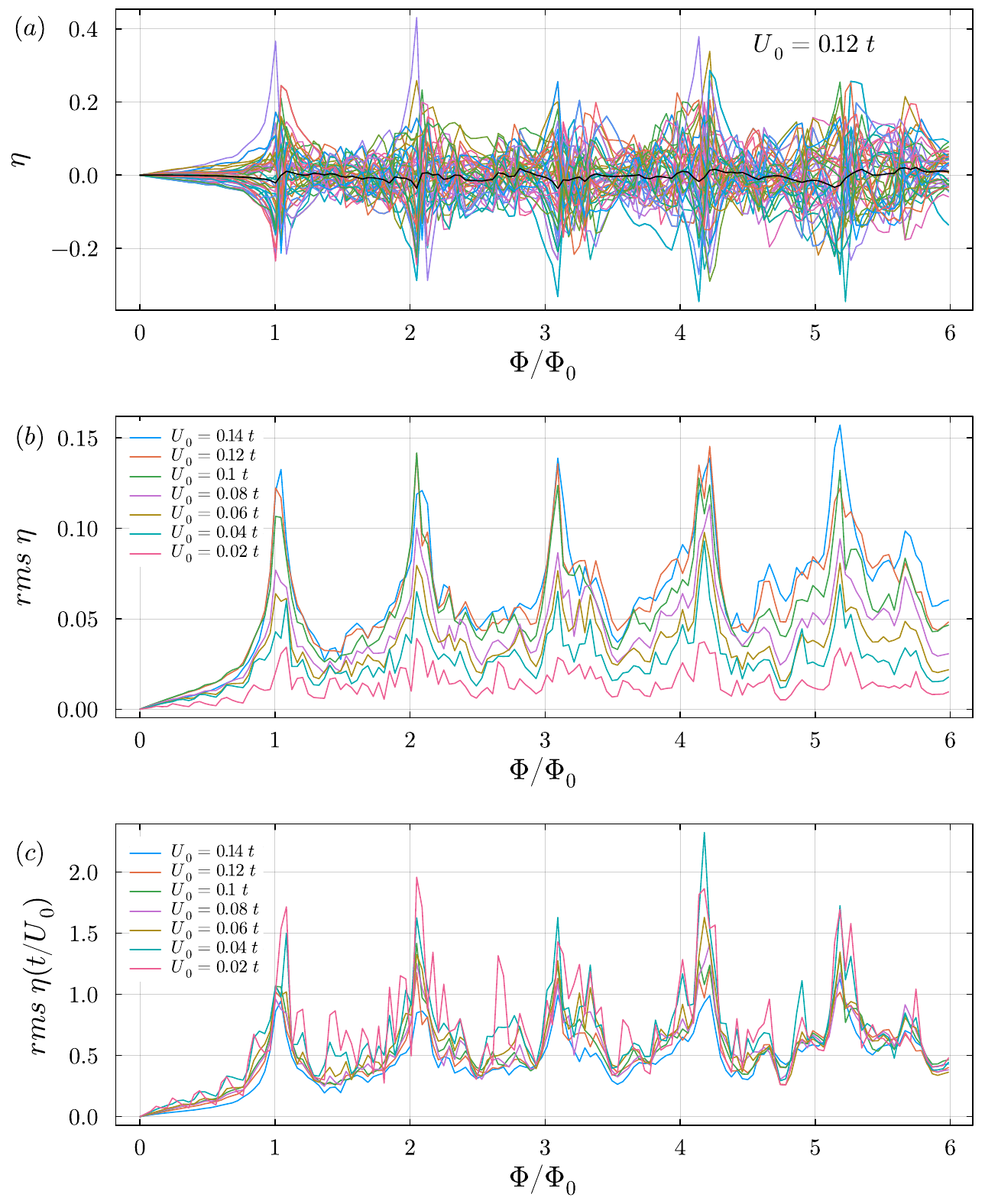}
	\caption{{\bf Statistical analysis of disordered configurations.} (a)  Rectification coefficient $\eta$ for 50 disorder configurations and a given value of the strength $U_0=1.2~t$. In black is the average rectification of the 50 random disorder configurations. (b) Root mean square of the rectification coefficient, ${\rm rms}~\eta\equiv \sqrt{\langle\eta^2\rangle-\langle\eta\rangle^2}$,  averaged over 50 different pseudorandom disorder configurations, versus the applied flux for different values of $U_0$. The curve corresponding to $U_0=1.2~t$ is the square root of the average of the square of the curves in (a). In (a) and (b) $\mu=0.3~t$, $L=20$, $W=100$ and there is no barrier. (c) ${\rm rms}~\eta\times(t/U_0)$ providing a scaling of the curves in (b) with the impurity strength. \label{Fig5}}     
\end{figure}

We assessed the fluctuations in the rectification coefficient by simulating random configurations of disorder of different strengths and as a function of the magnetic field. We assume no barrier, so that in absence of disorder the system responds with a given number of perfectly open channels. In Fig.~\ref{Fig5} (a) we show the magneto-fingerprint of 50 different disorder configurations for a disorder strength $U_0=0.12~t$, together with the statistical average over the 50 configurations in black. The latter gives a negligible rectification, as it is clearly understood considering that the rectification strongly varies with the disorder configurations, from positive to negative values.  In addition, we clearly see a fishbone structure in the fluctuations, with a central bone with fluctuations of a given size and spikes at values of the field corresponding to the nodes of the Fraunhofer pattern, with a period that slightly increases with the flux, in agreement with the clean case (not shown). In Fig.~\ref{Fig5}(b) we plot the root mean square of $\eta$ averaged over different disorder configurations. We clearly see that rms~$\eta$ starts from zero at zero field, as expected, and then saturates to a value that depends on the strength of disorder.  In addition, well-defined spikes appear at multiples of the flux quantum. After the first spike, corresponding to the first node of the Fraunhofer pattern, ${\rm rms}~\eta$ shows a field-independent value, that increases with the disorder strength. In Fig.~\ref{Fig5}(c), we show rms~$\eta$ rescaled by the impurity strength, rms~$\eta\times(t/U_0)$, and we see that the curves tend to fall on top of each other in the lobes of the Fraunhofer pattern but not on the nodes, although we point out that the number of configurations is very small for a meaningful statistics. In addition, we notice that the fluctuations in rms~$\eta$ for low value of $U_0$ also present regular jumps, as appreciated at low field, that originate from the activation/deactivation of the channels with the field. 

These signatures are clear manifestations of the fluctuations of the critical current due to disorder and magnetic field. At the same time, we recall that at finite magnetic field, the current cannot be expressed as a linear statistics in the transmission eigenvalues. A common trend emerges in Fig.~\ref{Fig5}(c), which points to a qualitatively universal behaviour that requires further study. We further notice that the spikes at the node of the Fraunhofer pattern are a clear manifestation of the mirror symmetry breaking and can help in distinguishing other intrinsic sources of JDE.

\begin{figure*}[t] 
\centering
\includegraphics[width=1.0\linewidth]{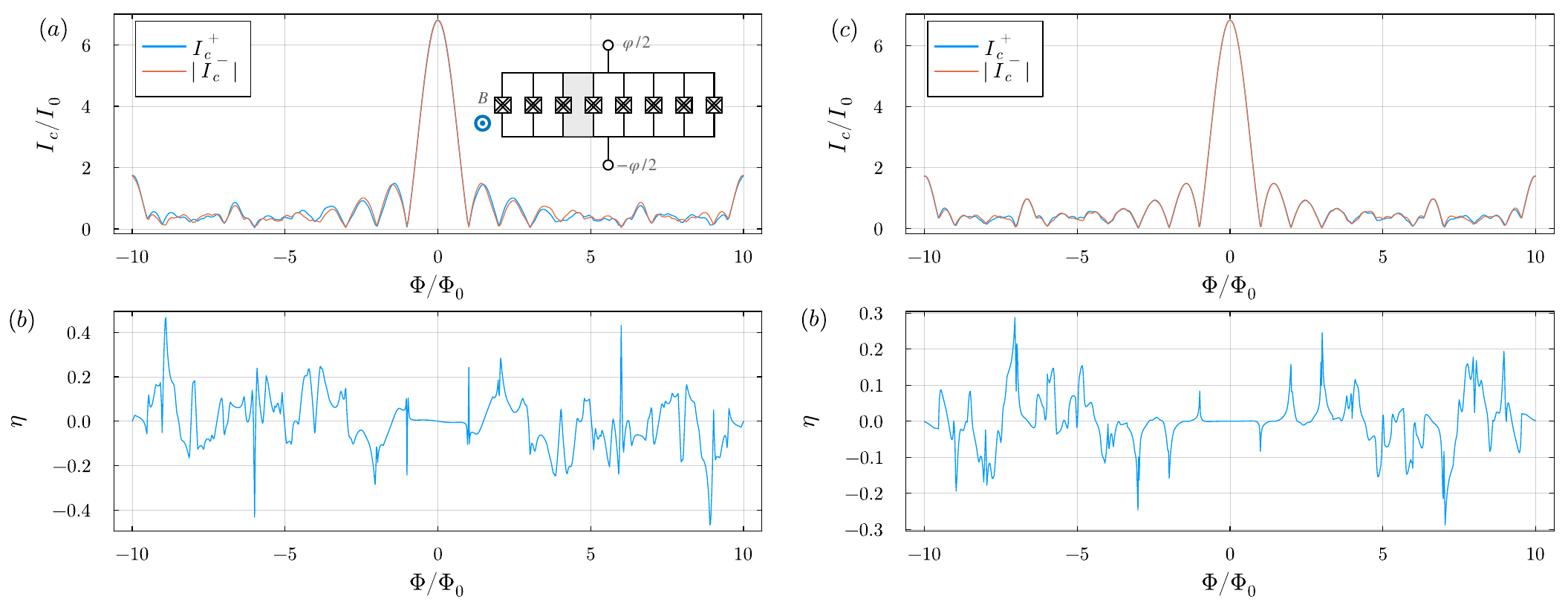}
\caption{\label{Fig6}{\bf JDE in the interference pattern of a set of Josephson junctions connected in parallel.} 
    Interference pattern of the critical currents $I^\pm_c(\Phi)$ of a multi-loop SQUID composed by $N=20$ junctions as a function of the total flux $\Phi$ through the junction, as arising from the energy-phase relation Eq.~\eqref{Eq:MultiLoopParallel}. (a) Equal area junctions with random transparencies  (shown in the inset). (b) Associated rectification coefficient. (c) Equal transmission $\tau=0.9$ junctions with random areas  (shown in the inset). (d) Associated rectification coefficient. }     
\end{figure*}

\subsection{Multiple-loop SQUID}
\label{Sec:Multi-loop-diode}

It is instructive at this point to consider a system that is expected to yield similar results, but that presents peculiar differences. This is the case of a system of $N$ Josephson junctions connected in parallel, each described by a single Energy-Phase Relations (EPR) of the form Eq.~(S9) of \footnote{Supplementary Material: Sec. III - Fraunhofer pattern of a multi-loop SQUID.} and enclosing $N-1$ loops threaded by a magnetic flux, that yields a Fraunhofer interference pattern. In Ref.~\cite{MertBorzkurt2023}, this system was engineered to give rise to a very high rectification diode effect by properly choosing the transmissions of the junctions and the phases in the loops. Here, we are interested in describing the system as a discretized version of the wide planar Josephson junction so far discussed and to study the effect of spatial inhomogeneity as a source of JDE. The total Josephson EPR describing the system reads
\begin{equation}\label{Eq:MultiLoopParallel}
E(\varphi)=-\sum_{j=1}^N E_j(\varphi-2\pi\Phi_{j}/\Phi_0),
\end{equation}
where $\Phi_{j}$ is the total magnetic flux enclosed in the loops before the $j$-th junction, and we choose the gauge in which the phase $\varphi$ drops at the first junction on the left side. When expressed in terms of Fourier components \cite{Note3}, the current reads
\begin{equation}\label{Eq:CPRgeneral}
    I(\varphi)=-iI_0\sum_{j=1}^N\sum_{n=1}^\infty n\epsilon_{n,j}e^{in\left(\varphi-\sum_{k<j}\varphi_{k}\right)}+{\rm c.c.},
\end{equation}
where $\epsilon_{n,j}$ are the Foruier component of the EPR of the $j$-th junction \cite{Note3}, $I_0=2\pi\Delta/\Phi_0$ and $\varphi_{j}=2\pi BA_j/\Phi_0$ is the  flux through the area $A_j$ of the loop between junctions $j-1$ and $j$, in units of $\Phi_0/2\pi$. The expression Eq.~\eqref{Eq:CPRgeneral} meets two fundamental conditions at the same time: {\it i)} it breaks time-reversal symmetry, and {\it ii)} it generally breaks mirror symmetry $j\to N+1-j$, thus allowing, in general, the appearance of a diode effect. This can be accomplished in two ways, {\it i)} by considering different transparencies $\tau_j$ and same areas of the loops, for which $\varphi_{j}=2\pi(j-1)\Phi/N\Phi_0$, or {\it ii)} by keeping the same transparency $\tau$ for the junctions and considering different areas $A_j$ of the loops, such that $\varphi_{j}=2\pi \sum_{i\leq j}A_jB/\Phi_0$. In addition, the transmissions $\tau_j$ need to be sizable. 

{\it i).} An example of a Fraunhofer interference pattern for a parallel of 20 junctions with the transmission randomly distributed between $0.85 < \tau_j < 0.95$ is shown in Fig.~\ref{Fig6}(a) (the inset shows the particular configuration of transmissions). A nonreciprocal Fraunhofer interference pattern appears, with rectifications up to 40 \% close to the nodes of the Fraunhofer pattern, as shown in Fig.~\ref{Fig6}(b) and in agreement with the planar junction case.

{\it ii).} The results for the case of slightly random fluxes $\varphi_{j}$ is shown in Fig.~\ref{Fig6}(c), for the case of equal transparencies $\tau=0.9$. We assume that $A_j=A_{\rm tot}(1-\epsilon)/(N-1)+\delta A_j$, with $\sum_{j=1}^{N-1}\delta A_j=\epsilon A_{\rm tot}$, in a way that we can study different disorder configurations with the same total area $A_{\rm tot}$, and choose $\epsilon=0.1$. In Fig.~\ref{Fig6}(d), we see that the rectification can reach 20\% and, as in the case of random transparencies, the most significant rectification values are observed at the nodes of the Fraunhofer pattern.

\begin{figure*}[t]
	\centering
    \includegraphics[width=1.0\linewidth]{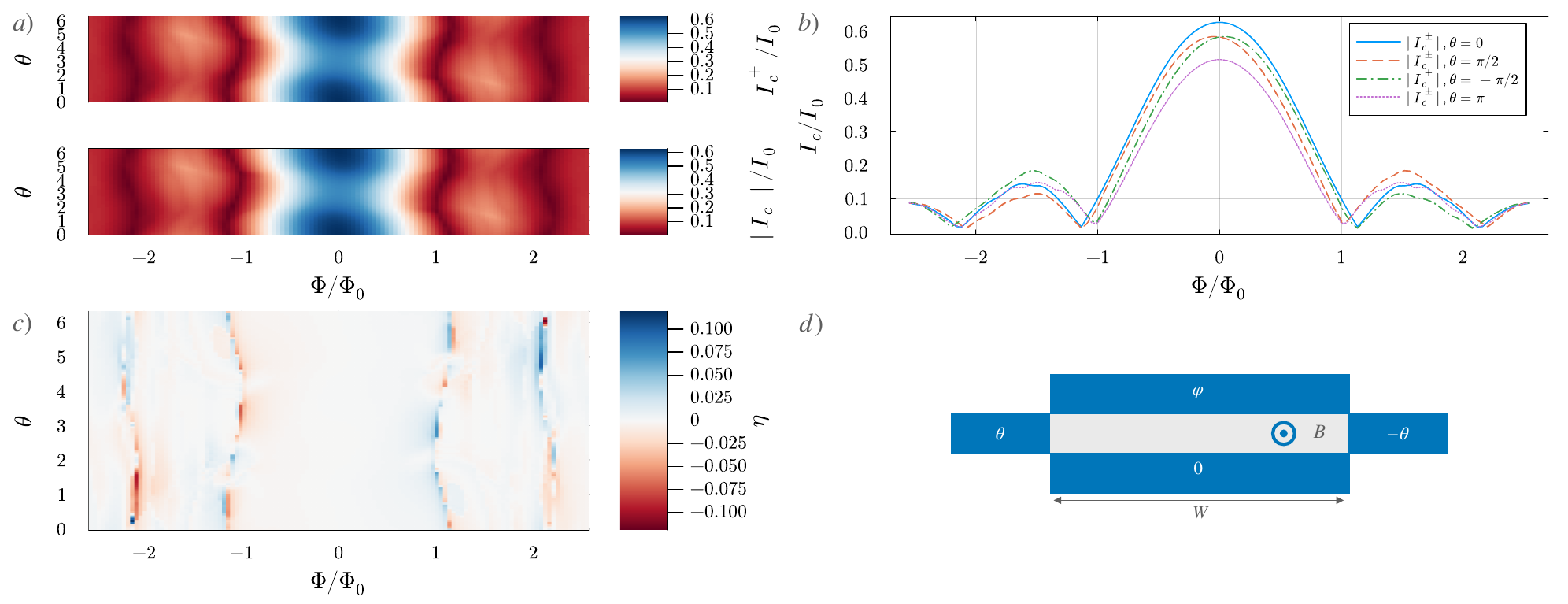}
	\caption{{\bf JDE in the Fraunhofer interference pattern of a four-terminal spatially symmetric junction.} (a) Critical currents $I_c^\pm$ as a function of the external flux $\Phi$ for the entire $(\theta,\Phi)$- dependence and (b) for four selected values of the phase $\theta=0,\pm\pi/2,\pi$. (c) Rectification coefficient $\eta$ for the critical currents shown in (a).  (d) Schematics of the junction, in which the bottom terminal serves as phase reference, and the top,  left, and right terminals are kept at phase $\varphi,\theta,-\theta$, respectively. The central normal region has width $W$, length $L$, pierced by a magnetic field $B$ that induces a flux $\Phi=B LW$.  
 \label{Fig7}}    
\end{figure*}

\subsection{Multi-terminal case}
\label{Sec:Multi-terminal}

The analysis presented for a two-terminal wide Josephson junction can be straightforwardly extended to the multiterminal case composed by $j=1,\ldots,N_l$ leads, each with phase $\varphi_j$ \cite{Note1}. This configuration has been recently realized in Ref.~\cite{coraiola2024}, where two of the three phase-differences of a four-terminal junction have been fixed and controlled by external fluxes in a two-loop configuration. Assuming a geometry of the scattering region and lead arrangements that yields a zero-field mirror symmetric multiterminal scattering matrix $s$, such that $M_x s(\Phi) M_x^{-1}=s(-\Phi)$, together with the Onsager reciprocity relations $s^T(\Phi)=s(-\Phi)$, it follows that the current in lead $j$ satisfies
\begin{equation}
  I_j(\boldsymbol{\varphi},\Phi)=-I_j(-{\cal P}\boldsymbol{\varphi},\Phi),  
\end{equation}
where ${\cal P}\boldsymbol\varphi$ is a permutation of the phases $\varphi_j$ that results from the mirror transformation. It then follows that a diode effect can appear already at zero field for values of the phases that are not invariant under mirror symmetry. In addition, a finite field yields a difference between the positive and negative critical current and a resulting diode effect in the Fraunhofer pattern for values of the phases that are not invariant under mirror symmetry.

To observe a Fraunhofer pattern, we consider a four-terminal symmetric structure, schematized in Fig.~\ref{Fig7}(d), with a wide bottom contact kept at phase zero for reference, a top contact kept at phase $\varphi$ with respect to the bottom terminal, and the lateral terminals kept at phase difference $\pm \theta$ with respect to the reference terminal. In a mirror-symmetric setup, the critical currents between the top and bottom terminals satisfy
\begin{equation}\label{Eq:Icp-4terminals}
  I_c^+(\theta,-\theta,\Phi)=-I_c^-(-\theta,\theta,\Phi),  
\end{equation}
where the mirror symmetry transformation maps $\theta\to-\theta$. This is seen in Fig.~\ref{Fig7} (a), where we show the critical currents for the entire range $0\leq \theta<2\pi$. In addition, in Fig.~\ref{Fig7} (b) we plot the critical currents $I_c^\pm$ four selected phase values $\theta=0,\pm \pi/2,\pi$. As expected, for $\theta=0$ the two critical currents perfectly overlap. Care has to be taken in choosing a symmetric gauge for the vector potential, in that the lateral superconducting contacts are sensitive to an overall constant shift of the vector potential.  For $\theta=\pi$ we see that $I_c^+(\pi,-\pi,\Phi)=-I_c^-(-\pi,\pi,-\Phi)$ and more generally we observe the behavior dictated by Eq.~\ref{Eq:Icp-4terminals}, as can be checked also in the entire $(\theta,\Phi)$-dependence of the rectification coefficient shown in Fig.~\ref{Fig7}(c). This setup shows how the phases of the lateral superconducting contacts can be used as knobs to control the diode effect also in a mirror-symmetric system. 

\section{Discussion}
\label{Sec:conculsions}

The presented analysis sheds light on the ubiquity of the diode effect in wide planar Josephson junctions, showing how seemingly different spatially nonuniform junctions yield qualitatively similar JDE. The spatial dependence of the phase that Cooper pairs acquire in an orbital magnetic field fully emerges in the Fraunhofer interference pattern in planar Josephson junctions, which turn out to be particularly suitable to highlight the relevant symmetries enabling the onset of JDE. In particular, the conditions of time-reversal and inversion symmetry breaking are known to be {\it necessary} conditions that do not guarantee the onset of a diode effect. In turn, casting the analysis in terms of mirror symmetry breaking is more stringent. Although it is only possible to demonstrate that junctions that do not break the relevant mirror symmetry cannot show JDE, all the cases studied do indeed show JDE. 

The relevant mirror plane is defined by the cross product between the orbital magnetic field and the direction of the phase bias (see Fig.~\ref{Fig0}). Interestingly, the same occurs in junctions with Rashba spin-orbit interactions and the Zeeman field in the plane\cite{baumgartner2022,costa2023,nadeem2023}. Indeed, the cross product between the direction of the phase bias and the Zeeman field defines the plane of the junction as the relevant one for the mirror symmetry. The latter is necessarily broken when a Rashba spin-orbit interaction is present in the system. 

We consider several cases of junctions that break the relevant mirror symmetry already in absence of magnetic field, either through smooth potentials compatible with back or side gates, or asymmetric junction shape, or through short-range scattering centers, such as those describing disorder at the atomic scale. All these conditions generally apply to physical devices, making the present analysis highly relevant to experiments.

Our symmetry-based results can be compared with other systems where JDE has been proposed, such as interacting quantum dots \cite{debnath2024,debnath2025}, systems hosting helical phases \cite{he2022,daido2022,ilic2022,peihao2024}, multiband superconductors \cite{yerin2024}, vortex-phase textures \cite{fukaya2024}, and through magnetization gradients \cite{roig2024}. In all these systems, the relevant mirror symmetry is broken.

The study of disorder averaging highlights interesting features of the rectification coefficient that can be ascribed to mesoscopic critical current fluctuations. In particular, the root mean square of the rectification appears to scale with the impurity strength, apart from the nodal points of the Fraunhofer pattern, where a strong magnification of the rectification occurs. The analysis presented points to the JDE as an interesting tool for studying mesoscopic critical current fluctuations, that calls for further study. 

Finally, we extend the result to a four-terminal device, showing how a phase difference on the control leads yields rectification for finite external magnetic fields, providing an additional knob for tuning the diode effect.

\section{Acknowledgments}

We acknowledge S. Heun for valuable discussions and for carefully reading the manuscript. L.C. acknowledges the Fondazione Cariplo under the grant 2023-2594. M.C. and F.G. acknowledge the EU’s Horizon 2020 Research and Innovation Framework Programme under Grants No. 964398 (SUPERGATE) and the PNRR MUR project PE0000023-NQSTI for partial financial support. F.G. acknowledges partial financial support under Grant No. 101057977 (SPECTRUM). E.S. acknowledges the project HelicS, project number DFM.AD002.206. We acknowledge NextGenerationEU PRIN project 2022A8CJP3 (GAMESQUAD) for partial financial support.

\section{Author contribution}

L.C. entirely performed the theoretical analysis, from conceiving the problem to the numerical simulations, A.C., E.S. and M.C. contributed to the discussions, the theoretical analysis, the experimental contextualization, and the writing of the paper, A.G., L.A. and F.G. contributed to the discussions and the reading of the work.

\section{Correspondence and Data Availability}
The data of the simulations are available upon request and correspondence and requests for material should be addressed to Luca Chirolli.

\clearpage
\onecolumngrid

\begingroup
\leftskip=0cm plus 0.5fil
\rightskip=0cm plus -0.5fil
\parfillskip=0cm plus 1fil
    \textbf{\large Supplemental Material for:} \\
    \textbf{\large ``Diode effect in the Fraunhofer pattern of disordered planar 
Josephson junctions''}
   \par
\endgroup
\vspace{0.5cm}

\setcounter{equation}{0}
\setcounter{figure}{0}
\setcounter{table}{0}
\setcounter{page}{1}
\makeatletter
\renewcommand{\theequation}{S\arabic{equation}}
\renewcommand{\thefigure}{S\arabic{figure}}
\twocolumngrid

\section{Scattering matrix formulation}
\label{SupplementaryNote-1}

We start by setting the framework for studying the diode effect in general Josephson junction systems. The Josephson current in a multiterminal setup composed of $N_l$ leads can be described using the scattering matrix formalism \cite{beenakker1991,beenakker2004,vanheck2014}, where the normal central region acts as a scattering region that connects external superconducting leads. The superconducting gap in the leads induces Andreev reflection, which converts the carriers at energy $\epsilon$ from particle-like to hole-like character and adds a phase shift $e^{-i{\rm arccos}(\epsilon/\Delta)}e^{i\varphi_j}$, with $\Delta e^{i\varphi_j}$ the superconducting gap in the leads $j$, and we assume all leads to have the same magnitude of the superconducting gap $\Delta$. We can define $N_l-1$ independent phase differences $\varphi_j$, with $j=1,\ldots, N_l-1$, with respect to the phase of a reference lead $j=0$ that is set to zero. The equilibrium currents in lead $j$, which can be expressed utilizing the free energy of the system  $F(\boldsymbol{\varphi})$ at temperature $T$ \cite{bardeen1969},
\begin{equation}
    I_j(\boldsymbol{\varphi})=\frac{2\pi}{\Phi_0}\frac{\partial F(\boldsymbol{\varphi})}{\partial \varphi_j},
\end{equation}
yielding $N_l-1$ independent currents, $I_j\equiv I_{0,j}\equiv I_j-I_0$, where the currents between any pairs of terminals is $I_{i,j}=I_{i}-I_j$.  If we apply a bias current between two given terminals and keep the phases on the other terminals fixed, we can define critical currents as
\begin{eqnarray}
    I_{c,j}^+=\max_{\varphi_j} I_j(\boldsymbol{\varphi}),\qquad
    I_{c,j}^-=\min_{\varphi_j} I_j(\boldsymbol{\varphi}),
\end{eqnarray}
and an associated rectification coefficient
\begin{equation}
    \eta_j=\frac{I_{c,j}^+-|I_{c,j}^-|}{I_{c,j}^++|I_{c,j}^-|}.
\end{equation}
Each terminal $j$ has $n_j$ open channels, and transport along these channels and through the central region is described by a unitary scattering matrix $s$ that has a  $N_l\times N_l$ block structure, with the block $s_{i,j}$ of dimensions $n_i\times n_j$. Unitarity $s^\dag s=1$ ensures overall current conservation. The Andreev spectrum is obtained by solving the eigenproblem \cite{beenakker1991,beenakker1997,beenakker2004}
\begin{equation}
s_{\rm A}s_{\rm N}\psi=\psi,
\end{equation}
where the normal (N) and Andreev (A) scattering matrices are given by
\begin{equation}
s_{\rm N}=\left(\begin{array}{cc}
s(\epsilon) & 0\\
0 & s^*(-\epsilon)
\end{array}\right),
\qquad 
s_{\rm A}=\alpha \left(\begin{array}{cc}
0 & r_A^*\\
r_A & 0
\end{array}\right),
\end{equation}
where  $r_A={\rm diag}
(e^{i\varphi_1}\openone_1,\ldots,e^{i\varphi_M}\openone_M)$, with $\openone_i$ a identity matrix with dimension given by the number of open channels in terminal $i$, and $\alpha=e^{-i~{\rm arccos}(\epsilon/\Delta)}$. 

In the Andreev approximation of a scattering region of linear size much smaller than the coherence length $\xi=\hbar v_F/\Delta$, we can neglect the energy dependence of the scattering matrix $s$, so that $s(\epsilon)=s(-\epsilon)\equiv s$. Following Ref.~\cite{vanheck2014}, the Andreev spectrum is therefore determined by
\begin{equation}
\left(\begin{array}{cc}
0 & s^\dag r^*_{\rm A}\\
s^Tr_{\rm A} & 0
\end{array}\right)\psi=\alpha\psi.
\end{equation}
We can transform the eigenvalue problem $X\psi=\alpha\psi$ in a way that $\frac{1}{2}(X+X^{-1})\psi=\frac{1}{2}(\alpha+\alpha^{-1})\psi=\frac{\epsilon}{\Delta}\psi$, so that the general Andreev spectrum is obtained by solving the secular equation \cite{vanheck2014} 
\begin{equation}\label{Eq:EigA}
\left(\begin{array}{cc}
0 & A\\
A^\dag & 0
\end{array}\right)\psi=\frac{\epsilon}{\Delta}\psi,
\end{equation}
where the complex symmetric matrix $A$ is given by
\begin{equation}
A=\frac{1}{2}(r_{\rm A}s+s^Tr_{\rm A}).
\end{equation}
By construction, the eigenproblem Eq.~(\ref{Eq:EigA}) is particle-hole symmetric, and the singular values of $A$ uniquely determine the spectrum. In the presence of time-reversal symmetry, the scattering matrix is symmetric $s^T=s$. 

For a two-terminal device, it is possible to express the scattering matrix through the polar decomposition \cite{beenakker1997,vanheck2014}. It follows that the Andreev spectrum acquires the general energy-phase relation (EPR) form $E^\pm(\varphi)=\pm \sum_pE_p(\varphi)$, with 
\begin{equation}\label{Eq:EPR}
E_p(\varphi)=\Delta\sqrt{1-\tau_p\sin^2(\varphi/2)},
\end{equation}
where the sum is extended over the open channels of transmission $\tau_p$. The free energy is then obtained by populating the levels according to a Fermi-Dirac distribution centered at zero energy.

In the presence of an orbital magnetic field, it is not possible to establish a relation between the energy and the phase bias involving the zero-field channels' transmission probabilities $\tau_p$.

\subsection{Action of symmetry transformations}

\subsubsection{Mirror $M_x$}

For the case of a scattering matrix $s$ that in absence of the external magnetic field is mirror symmetric, such that it transforms as $M_x s(\Phi)M_x^{-1}=s(-\Phi)$,  we have
\begin{eqnarray}
    M_xA(\varphi,\Phi)M_x^{-1}&=&M_x\frac{1}{2}( r_A s(\Phi)+s^T(\Phi)r_A)M_x^{-1}\nonumber\\
    &=&\frac{1}{2}(r_A s(-\Phi)+s^T(-\Phi)r_A)\nonumber\\
    &=&r_A\frac{1}{2}( s(-\Phi)r_A^*+r_A^*s^T(-\Phi))r_A\nonumber\\
    &=&r_A\frac{1}{2}( s^T(\Phi)r_A^*+r_A^*s(\Phi))r_A\nonumber\\
    &=&r_A A(-\varphi,\Phi) r_A
\end{eqnarray}
where we used that $M_x r_AM_x^{-1}=r_A$, $r_A(-\varphi)=r_A^*(\varphi)=r_A^{-1}(\varphi)$. More generally, if a scattering matrix transforms under the mirror operation as $M_x s(\Phi)M_x^{-1}=Gs(-\Phi)G^\dag$, with $G$ a field-independent unitary matrix, the singular values of $A(\varphi,\Phi)$ and $A(-\varphi,\Phi)$ are the same and a JDE can be ruled out.

\subsubsection{Mirror $M_y$}

For the case of symmetry about $M_y$, such that $M_ys(\Phi)M_y^{-1}=s(-\Phi)$, we have that
\begin{eqnarray}
M_yA(\varphi,\Phi)M_y^{-1}&=&M_y\frac{1}{2}( r_A s(\Phi)+s^T(\Phi)r_A)M_y^{-1}\nonumber\\
    &=&\frac{1}{2}(r_A^* s(-\Phi)+s^T(-\Phi)r_A^*)\\
    &=&\frac{1}{2}(r_A^*s^T(\Phi)+s(\Phi)r_A^*)\nonumber\\
    &=&r_A^*\frac{1}{2}( s^T(\Phi)r_A+r_As(\Phi))r_A^*\nonumber\\
    &=&r^*_A A(\varphi,\Phi) r_A^*,
\end{eqnarray}
where we have used that $M_y r_A M_y^{-1}=r^*_A$. In this case, we cannot say anything concerning a mirror symmetry about the $y$ axis, and the statement is non-conclusive.

\subsubsection{Inversion symmetry}

In the presence of inversion symmetry, we have
\begin{equation}
    I s(\Phi) I^{-1} = s(\Phi).
\end{equation}
We can use the results for the mirror symmetry by noticing that
\begin{equation}
    \hat{I}=M_x M_y M_z,
\end{equation}
and that the magnetic field is symmetric under mirror symmetry about any plane orthogonal to it. We also assume that, since the spin plays no role and that the system is 2D, the mirror about the $z$ direction acts trivially. In this case, choosing the magnetic field about the $z$ direction, we have that
\begin{eqnarray}
    \hat{I}A(\varphi,\Phi)\hat{I}^{-1}&=&M_x M_y M_z A(\varphi,\Phi) M_z^{-1}M_y^{-1}M_x^{-1}\nonumber\\
    &=&M_x M_y A(\varphi,\Phi)M_y^{-1}M_x^{-1}\nonumber\\
    &=&M_x r_A^* A(\varphi,\Phi) r_A^* M_x^{-1}\nonumber\\
    &=&r_A^* r_A A(-\varphi,\Phi) r_A r_A^*\nonumber\\
    &=& A(-\varphi,\Phi),
\end{eqnarray}
so in the presence of inversion symmetry, we can rule out a diode effect.

\begin{figure}[t]
	\centering
    \includegraphics[width=1.0\linewidth]{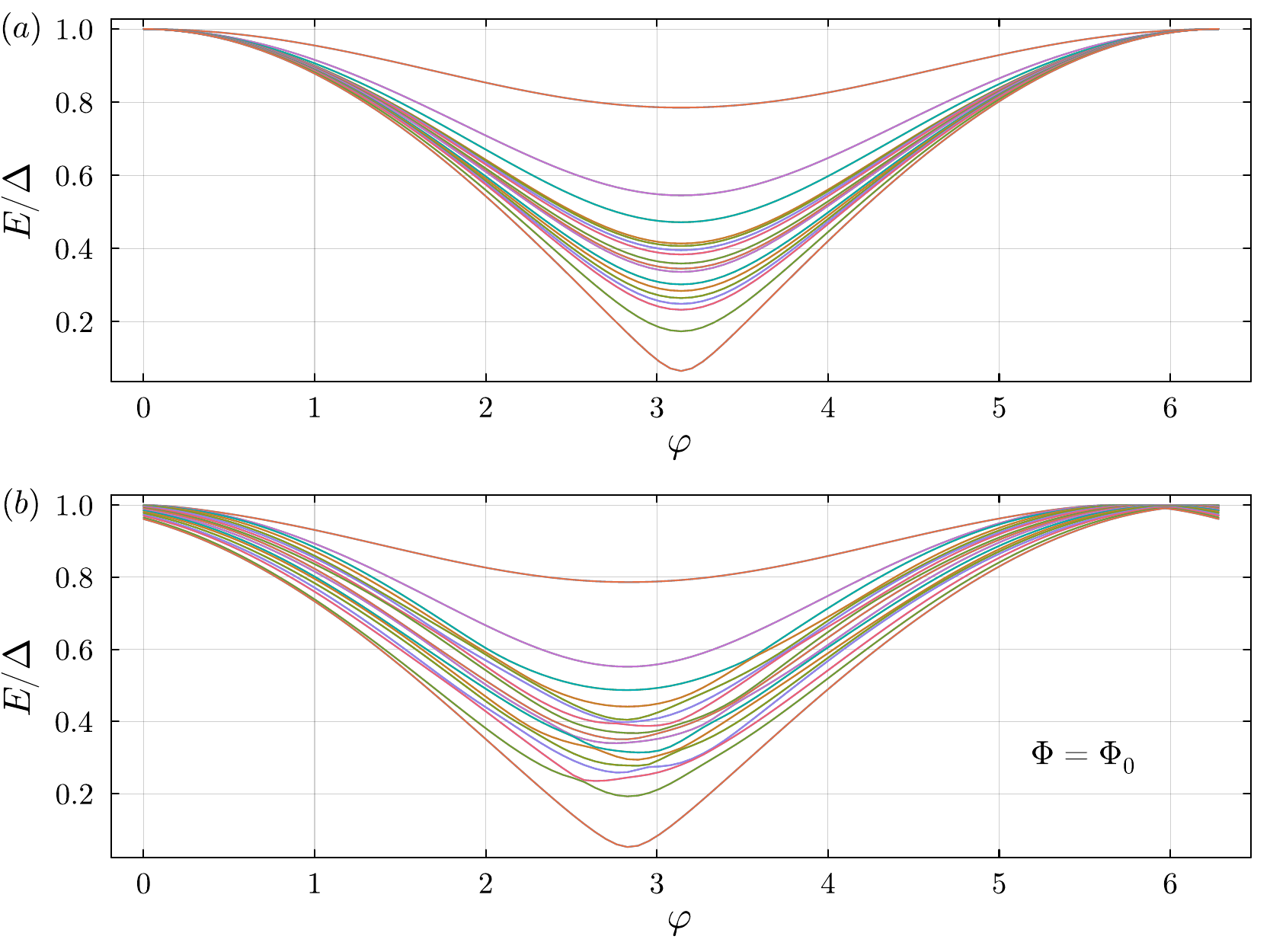}
	\caption{{\bf Andreev spectrum of a disordered junction:} (a) in zero magnetic field and (b) with $\Phi=\Phi_0$. The local impurity strength is set to $-U_0/2<\delta U_i<U_0/2$, with $U_0=0.1~t$, and barriers with $U_b=0.2~t$ are added at the interface with the superconducting leads. The central scattering area is a rectangle with lateral size $L=20$, $W=100$, and the chemical potential is set to $\mu=0.3~t$. \label{Fig8}}     
\end{figure}

\section{Comparison with realistic systems}
\label{App:detailsSimulation}

The energy scale of the tight-binding model is the hopping amplitude $t$ and the associated bandwidth is $8t$, as it follows from the dispersion $\epsilon_{\bf k}=4t-\mu-2t \cos(a k_x)-2t\cos(a k_y)$. By expanding the dispersion around zero momentum we obtain a parabolic dispersion $\epsilon_{\bf k}\sim ta^2(k_x^2+k_y^2)-\mu$ and comparing it with a parabolic dispersion $\epsilon_{\bf k}=\hbar^2(k^2_x+k_y^2)/2m^*-E_F$ we estimate $a=\hbar/\sqrt{2tm^*}$ and $E_F=\mu$. Assuming an effective mass $m^*=0.02 m$, a chemical potential $\mu=0.3~t$, and an energy scale $t=0.2~{\rm eV}$, we obtain an effective lattice constant $a=3~{\rm nm}$, so that the device we simulate is about $W=300~{\rm nm}$ large and $L=60~{\rm nm}$ long. A Fermi energy on order of $E_F\sim 0.06~ {\rm eV}$ (for hole carriers) is much larger than typical superconducting gaps, for example of Al, that is on order of $0.163~{\rm meV}$ at zero temperature, and $2.32~{\rm meV}$ for Nb. A relevant scale is provided by $E_c=\hbar/\tau_{\rm dw}$, where $\tau_{\rm dw}=\sqrt{WL}/v_F$ is the dwelling time, with $v_F=\hbar k_F/m^*$ the Fermi velocity, yielding an energy scale $E_c=\hbar/\tau\sim 5~{\rm meV}$. For a superconducting gap smaller than the energy scale $E_c$, the energy dependence of the scattering matrix can be neglected and the system falls in the regime in which the coherence length is larger than the linear size of the system.

The Fermi wavelength is $\lambda_F=2\pi \hbar/\sqrt{2m^*E_F}=2\pi a/\sqrt{E_F/t}$, that is comparable with the lattice spacing, thus yielding more a metallic behaviour. A proper modelling of a semiconductor would require a much smaller Fermi energy, with consequent much longer Fermi wavelength, that would also require a larger size of the system to accurately describe many channels. At the same time, for clean and weakly disordered systems, the parameters used are equally good for describing a semiconductor. Within this approximation the simulation describes a system that is perhaps a factor of 5-10 smaller than the typical planar Josephson junctions currently realized in labs, but in a regime of high carrier density.

For the system with local short-wavelength disorder, the variation of the local potential $\delta U_i$ is within the window $-U_0/2<\delta U_i<U_0/2$, with $U_0=0.1 t$. This is a relatively weak variation, yielding a mean free path $\ell_{\rm mf}$ on order of the system size, although it is hard to estimate it precisely. The junction then falls in the regime $\ell_{\rm mf},\sqrt{LW} < \xi$, that is disordered ballistic regime \cite{beenakker1991}.

\section{Fraunhofer pattern of a multi-loop SQUID}
\label{SupplementaryNote-3}

The EPR of Eq.~\eqref{Eq:EPR} describes a general Andreev bound state between two superconducting leads and across a high transparency normal region. It has a general functional form on the phase difference and can also arise in the case of a series of two Josephson junctions, as shown in Ref.~\cite{MertBorzkurt2023}, and for a diffusive system, know as Kulik Omel'yanchuk formula. 

For the case of an Andreev state, the current can also be calculated as a series expansion in the tunnel transparency by considering multiple Cooper pair transport across the junction through standard techniques of many-body theory \cite{mahan}. One of the characteristics of the EPR Eq.~\eqref{Eq:EPR} is the alternation in sign between the even and odd harmonics, which is rooted in the Andreev reflection phase "$i$" at zero energy. This way, a net Cooper pair is transferred between two superconducting leads through a barrier of transmission $\tau$ and via a process involving two Andreev reflections at the two superconducting leads, thus carrying a $\pi$ phase-shift. Transfer of multiple Cooper pairs carries in general a minus sign for an odd number and a plus sign for an even number.

\begin{figure}[t]
\includegraphics[width=1.0\linewidth]{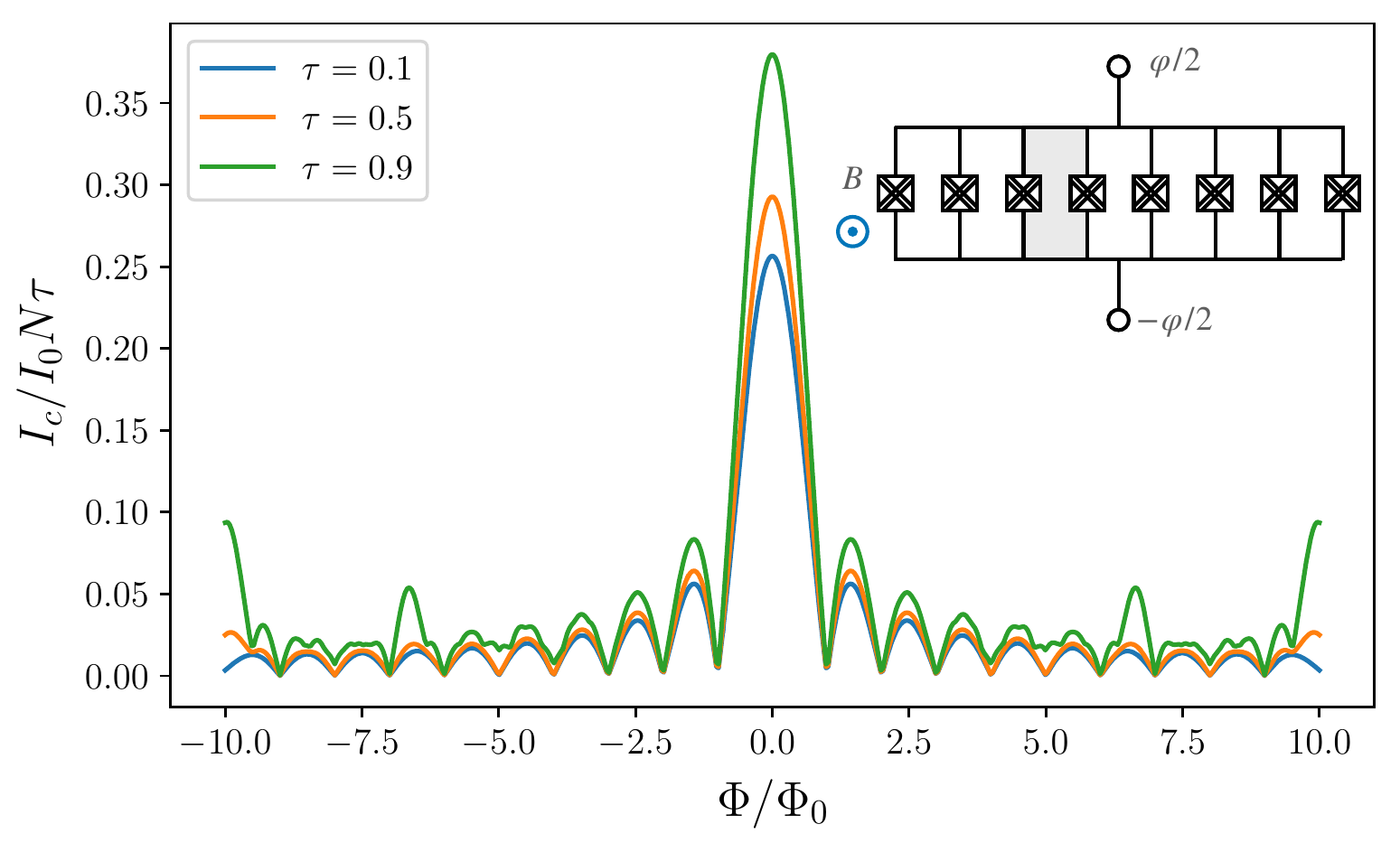}
    \caption{\label{Fig9}{\bf Fraunhofer interference pattern of the critical current $I_c(\Phi)$ of a multi-loop SQUID.} The system is composed by $N=20$ junctions, each described by an EPR as in Eq.~\eqref{Eq:EPR} and the junctions are connected in parallel, as shown in the inset. The interference pattern is shown as a function of the total flux $\Phi$ through the junction, for equal junctions of transparency $\tau=0.1,0.5,0.9$. Inset: schematics of a multi-loop SQUID formed by a parallel of several Josephson junctions connected to superconducting leads on the top and the bottom, kept at phase difference $\varphi$.}     
\end{figure}

A direct calculation of the Fourier component of the Andreev state EPR confirms the sign alternation of even and odd harmonics and reads \cite{willsch2024} 
\begin{equation}\label{Eq:EPR-Fourier}
E(\varphi)/\Delta=-\epsilon_0-2\sum_{n=1}^\infty \epsilon_n\cos(n\varphi),
\end{equation}
with the coefficients $\epsilon_n$ given by
\begin{equation}
\epsilon_n=\frac{\tau^n}{4^n}\frac{\Gamma(3/2)}{n!\Gamma(3/2-n)}{}_2F_1(n-1/2,n+1/2;1+2n,\tau),    
\end{equation}
where $\Gamma(z)$ is the Gamma function and ${}_2F_1(a,b;c;z)$ is the hypergeometric function \cite{gradshteyn}. The coefficients $\epsilon_n=(-1)^{n+1}|\epsilon_n|$ are positive for odd $n$ and negative for even $n$.

For a mirror symmetric system of (single-channel) junctions of equal transparency $\tau$ and equal area loops threaded by a total flux $\Phi=(N-1)BA$, with $A$ the area of a single loop, it is easily shown that $I(\varphi,\Phi)=I(\varphi+2\pi(N-1)\Phi/(N\Phi_0),-\Phi)$. Recalling the definitions of the critical currents and the Onsager relations, it follows that no diode effect appears for any finite external field. In particular, the current reads
\begin{eqnarray}
    I(\varphi)&=&2I_0\sum_{n=1}^\infty n\epsilon_n\frac{\sin(\pi n N\Phi/(N-1)\Phi_0)}{\sin(\pi n\Phi/(N-1)\Phi_0)}
    \nonumber\\
    &\times&\sin(n(\varphi-\pi\Phi/\Phi_0)).
\end{eqnarray}
The dependence on the external flux of the critical current is shown in Fig.~\ref{Fig9}, where a Fraunhofer interference pattern appears that is periodic with period $\Phi=(N-1)\Phi_0$, i.e., a flux quantum $\Phi_0$ threading each unit cell (a gray area in the inset of Fig.~\ref{Fig9}). For low transparency $\tau$ the critical current reads
\begin{equation}
    I_c(\Phi)=\frac{\tau I_0}{4}\left|\frac{\sin(\pi N\Phi/(N-1)\Phi_0)}{\sin(\pi \Phi/(N-1)\Phi_0)}\right|,
\end{equation}
and exact zeros of the interference pattern appear at fractional values of the flux quantum, $\Phi=k(N-1)\Phi_0/N$, with $k=1,\ldots,N-1$. For large $N$ we recover the typical expression of the Fraunhofer interference pattern, $I_c(\Phi) = \frac{\tau I_0 (N-1)}{4}|{\rm sinc}(\pi\Phi/\Phi_0)|$, with ${\rm sinc}(x)=\sin(x)/x$ (the factor 1/4 comes from the first order expansion in $\tau$ of Eq.~\eqref{Eq:EPR}) \cite{tinkham2004introduction}, with zeros appearing at integer values of the flux quantum, $\Phi=k\Phi_0$. 

For general transparency $\tau$, a closed expression for the critical current cannot be derived, and we have to resort to numerics. No exact zero appears in the Fraunhofer pattern; however, for fractional values of the flux $\Phi=k(N-1)\Phi_0/N$, the critical current generally displays minima. Indeed, the amplitudes of the current $A_n(k)=\sin(\pi nk)/\sin(\pi nk/N)$ are zero unless for $kn$ multiple of $N$. For $N$ even, the critical current has a secondary maximum at $k=N/2$ that involves only even $n$ harmonics, so it is on the order of $\tau^2$. For $N$ prime number, the current at the minima has only harmonics that are multiples of $N$, so that it is on order $\tau^N$, and for general $N=pq$ the critical current at the minima is on order of $\tau^{\min(p,q)}$.

In the case of broken mirror symmetry, it is easily shown that for $\tau_j\ll 1$, such that only the first harmonic of the CPR is non-negligible, no JDE arises. Indeed, by introducing $Z(\Phi)\equiv\sum_{j=1}^N \epsilon_{1,j}e^{i\sum_{k<j}\varphi_{k}(\Phi)}$,  we can write
\begin{eqnarray}
    I(\varphi)&=& 2I_0|Z(\Phi)|\sin(\varphi-{\rm arg}(Z(\Phi)),
\end{eqnarray}
from which by noticing that $|Z(\Phi)|=|Z(-\Phi)|$ immediately follows that $I_c^+(\Phi)=-I_c^-(\Phi)$. 

{\it i)} For the case in which the fluxes through the loops are all equal but the junctions have different transparencies, by simple manipulations it can be shown that if the junctions are not inversion symmetric under the transformation $j\to N+1-j$, such that $\epsilon_{n,j}\neq \epsilon_{n,N+1-j}$, we have 
\begin{equation}
I(\varphi,\Phi)\neq I(\varphi+2\pi(N-1)\Phi/N\Phi_0,-\Phi),   
\end{equation}
so that we cannot rule out the JDE. 

{\it ii)} In the case of equal transparencies and different areas, we can cast the current in the form $ I(\varphi)=2I_0\sum_{n=1}^\infty n\epsilon_{n}|Z_n(\Phi)|\sin({\rm arg}(Z_n(\Phi))-n\varphi)$ where $Z_n(\Phi)=\sum_{j=1}^Ne^{-in\sum_{k<j}\varphi_{k}}$ and the irregularity of the phase pattern only implies that $I(\varphi,-\Phi)=-I(-\varphi,\Phi)$, without ruling out JDE.

\bibliography{JFD-bib}

\end{document}